\documentclass[twocolumn]{jpsj3} 

\title{Magnetically Robust Non-Fermi Liquid Behavior in Heavy Fermion Systems with f$^2$-Configuration: Competition between Crystalline-Electric-Field and Kondo-Yosida Singlets}

\author{Shinya \textsc{Nishiyama}\thanks{E-mail address: nishiyama@blade.mp.es.osaka-u.ac.jp}, Hiroyasu \textsc{Matsuura}, and Kazumasa \textsc{Miyake} }
\inst{Division of Materials Physics, Department of Materials Engineering Science, Graduate School of Engineering Science, Osaka University, Toyonaka, Osaka 560-8531, Japan}

\recdate{\today}
\abst{In f$^2$-based heavy fermion systems with a crystalline-electric-field (CEF) singlet ground state, the Non-Fermi Liquid (NFL) arises around the quantum critical point (QCP) due to the competition between the CEF singlet and the Kondo-Yosida singlet states.
In such a case, the characteristic temperature $T_{\rm F}^{*}$ at which the entropy starts to decrease toward zero is suppressed by the effect of the competition, compared to both energy scales characterizing each singlet state, the lower Kondo temperature ($T_{\rm K2}$) and the CEF splitting ($\Delta$).
We show that in the case of tetragonal symmetry $T_{\rm F}^{*}$ is not affected by the magnetic field up to $H_z^{*}$ which is determined by the distance from the QCP or characteristic energy scales of each singlet states, not by $T_{\rm F}^{*}$ itself. 
As a result, in the vicinity of QCP, there are parameter regions where the NFL is robust against the magnetic field, at an observable temperature range $T > T_{\rm F}^{*}$, up to $H_z^{*}$ which is far larger than $T_{\rm F}^{*}$ and less than $\min(T_{\rm K2},\,\Delta)$.
Our result suggests that such an anomalous NFL behavior can arise also in systems with other CEF symmetry, which might provide us with the basis to understand the anomalous behaviors of UBe$_{13}$.  
}
\kword{numerical renormalization group, heavy fermion, crystalline-electric-field singlet, f$^2$, non-Fermi liquid, Kondo effect, magnetic field effect}

\begin{document}
\maketitle

\section{Introduction}
In the last decade or so, non-Fermi liquid (NFL) behaviors around quantum critical point (QCP) have been one of main issues in physics, not only in heavy fermion systems\cite{lohneysen}, but also in those exhibiting the Mott transition\cite{imada}.
Of these NFL behaviors, those of heavy fermion systems with f$^2$-configuration form a kind of subclass in which the QCP is triggered by local criticalities: such as the two-channel Kondo effect (TCKE) due to the non-Kramers doublet state\cite{cox,cox1}, and that caused by the competition between the crystalline-electric field (CEF) singlet and the Kondo-Yosida (K-Y) singlet states\cite{yotsuhashi,hattori}.
The former TCKE was reported to be observed in La$_{1-x}$Pr$_x$Pb$_3$ that has a $\Gamma_3$ non-Kramers doublet ground state in the cubic symmetry\cite{kawae}.
The NFL behaviors in Th$_{1-x}$U$_x$Ru$_2$Si$_2$ were understood in a unified way by assuming that the system is located near the phase boundary between the CEF singlet and the K-Y singlet states\cite{yotsuhashi}.
owever, a detailed study about the magnetic field dependence on NFL behaviors has not been performed so far. \par
In the present paper, we investigate the magnetic field dependence of NFL behaviors in the specific heat $C_{\rm imp}(T)$ and the entropy $S_{\rm imp}(T)$ due to f-electrons with the two-orbital impurity Anderson model in a tetragonal symmetry with the CEF singlet ground state on the basis of the numerical renormalization group (NRG) method\cite{wilson,krishna}.
We discuss how the magnetic field, $H_z$, changes the characteristic temperature, $T_{\rm F}^{*}$, which is defined as the temperature at which the temperature derivative of entropy, $\partial S_{\rm imp}(T)/\partial (\log\,T)$, takes the maximum value as $S_{\rm imp}(T)$ approaching 0 as $T \rightarrow 0$.
In the vicinity of the QCP, $T_{\rm F}^{*}$ is suppressed by the effect of the competition between the CEF singlet and the K-Y singlet states for $H_z=0$, and the NFL behaviors occur at $T_{\rm F}^{*} < T < T_{\rm K2}$, where $T_{\rm K2}$ is the lower Kondo temperature of two orbitals, as in the case of TCKE.
The magnetic field is shown not to affect $T_{\rm F}^{*}$ up to a certain value $H_z^{*}$ which is determined approximately by the condition that the effect of the magnetic field, destroying a criticality of the TCKE type, becomes comparable to the effect of the deviation from the criticality at $H_z=0$.
$H_z^{*}$ so determined is far larger than $T_{\rm F}^{*}(H_z=0)$ for a reasonable set of parameters.
As a result, the NFL behaviors become robust against the magnetic field up to $H_z^{*} \sim T_{\rm K2}$ which is about hundred times larger than $T_{\rm F}^{*}(H_z=0)$.\par
This paper is organized as follows.
In \S 2, the model Hamiltonian is introduced and transformed into a form suitable for the NRG calculation.
In \S 3, we discuss how the characteristic temperature $T_{\rm F}^{*}$ is affected by the effect of the competition between the CEF singlet and the K-Y singlet states in the case of $H_z =0$.
In \S 4, we demonstrate the magnetic field dependence of $T_{\rm F}^{*}$ and $\gamma_{\rm imp}(T) = C_{\rm imp}/T$.
In the vicinity of the QCP, there are parameter regions where $-\log\,T$ behavior of $\gamma_{\rm imp}$, at temperature $T_{\rm F}^{*} < T < {\rm min}(T_{\rm K},\Delta)$, is robust against the magnetic field.
In \S 5, we investigate how such an anomalous NFL is affected by the change of the characteristic energy scale of two singlet states. 
In \S 6, we summarize our results and discuss their applicability for understanding the magnetically robust NFL behaviors observed in UBe$_{13}$ because such an NFL being robust against the magnetic field can arise in systems with other symmetry if the K-Y singlet state and the CEF singlet state compete for the ground state.

\section{Model Hamiltonian}
In this section, we recapitulate discussions of ref.\citen{yotsuhashi} about how to derive the model Hamiltonian for discussing the competition between the K-Y singlet and the CEF singlet states in f$^2$-configuration on the basis of the $j-j$ coupling scheme in the tetragonal symmetry.
We restrict the $f^1$ state within two low-lying doublet states out of three doublets of $j=5/2$ orbitals, and allot the pseudospin representation for these states as follows:
\begin{eqnarray}
	\label{2.1a}
	\vert \Gamma_{7+}^{(2)} \rangle &=& \frac{3}{\sqrt{14}} \vert  + \frac{5}{2} \rangle - \sqrt{\frac{5}{14}} \vert -\frac{3}{2} \rangle \equiv \vert \uparrow, 0 \rangle, \\
	\vert \Gamma_{7-}^{(2)} \rangle &=& -\frac{3}{\sqrt{14}} \vert -\frac{5}{2} \rangle + \sqrt{\frac{5}{14}} \vert +\frac{3}{2} \rangle \equiv \vert \downarrow, 0 \rangle, \\
	\vert \Gamma_{6,+} \rangle &=& \vert + \frac{1}{2} \rangle \equiv \vert 0, \uparrow \rangle, \\
	\label{2.1d}
	\vert \Gamma_{6,-} \rangle &=& \vert - \frac{1}{2} \rangle \equiv \vert 0, \downarrow \rangle. 
\end{eqnarray}
Here, for example, $\vert \hspace{-2.0mm}\uparrow,0 \rangle$ represents the state where orbital 1 ($\Gamma_{7}^{(2)}$) with up pseudospin is occupied and orbital 2 ($\Gamma_6$) is empty.
We also restrict the $f^2$ state within four low-lying states out of states allowed in $J=4$ manifold, and construct these four states with the direct product of $f^1$ states.
Here, we have discarded states where two f-electrons occupy the same orbital, $\vert \hspace{-1.0mm} \uparrow \downarrow, 0 \rangle$, $\vert 0, \uparrow \downarrow \rangle$, because the intra-orbital Coulomb repulsion is larger than the inter-orbital one.
Then, low-lying four $f^2$ states are expressed as 
\begin{eqnarray}
	\label{2.2a}
	\vert \Gamma_4 \rangle =&  \frac{1}{\sqrt{2}} \left( \vert  +2 \rangle - \vert -2 \rangle \right) =\frac{1}{\sqrt{2}} \left( \vert \hspace{-1.0mm} \downarrow, \uparrow \rangle - \vert \hspace{-1.0mm} \uparrow, \downarrow \rangle \right),\ \ \ \\
	\label{2.2b}
	\vert \Gamma_3 \rangle =& \frac{1}{\sqrt{2}} \left( \vert +2 \rangle + \vert -2 \rangle \right) =\frac{1}{\sqrt{2}} \left( \vert \hspace{-1.0mm} \uparrow, \downarrow \rangle + \vert \hspace{-1.0mm} \downarrow, \uparrow \rangle \right),\ \ \ \\
	\label{2.2c}
	\vert \Gamma_{5,+}^{(2)} \rangle =& \beta \vert +3 \rangle - \alpha \vert -1 \rangle = \vert \hspace{-1.0mm} \uparrow, \uparrow \rangle,\\ 
	\label{2.2d}
	\vert \Gamma_{5,-}^{(2)} \rangle =& \beta \vert -3 \rangle - \alpha \vert +1 \rangle = \vert \hspace{-1.0mm} \downarrow, \downarrow \rangle.
\end{eqnarray}
It is noted that we cannot determined coefficients, $\alpha$ and $\beta$, because we have discarded one of the doublet in f$^1$-configuration. 
Therefore, in this paper, we take its $j-j$ coupling representation as f$^2$ states with $\Gamma_5^{(2)}$ symmetry as shown in Appendix including the derivation of eqs.(\ref{2.1a})-(\ref{2.2d}).\par
\begin{figure}[t]
\begin{center}
	\includegraphics[width = 0.35\textwidth]{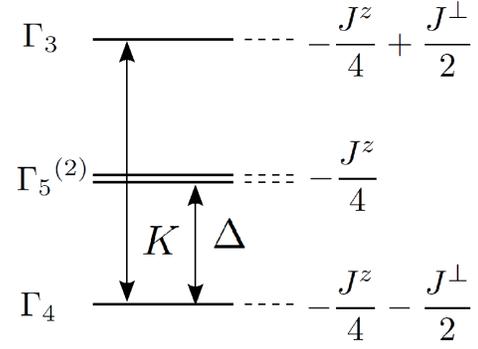}
	\caption{CEF level scheme of low-lying $f^2$ states and their eigenstates. }
	\label{fig1} 
\end{center}
\vspace{-8mm}
\end{figure}

We assume that the CEF ground state is the singlet ($\Gamma_4$), the first excited CEF states are magnetic doublet ($\Gamma_5$) with the excitation energy $\Delta$, and the second excited CEF state is the singlet ($\Gamma_3$) with the excitation energy $K$, as shown in Fig.\ref{fig1}.
Such a CEF level scheme can be reproduced by introducing the ``antiferromagnetic Hund's-rule coupling'' for the pseudospin as 
\begin{equation}
	\mathcal{H}_{\rm Hund} = \frac{J_{\perp}}{2} \left[ S_1^{+}S_{2}^{-} +S_1^{-}S_{2}^{+} \right] + J_z S_{1}^{z}S_{2}^{z},
	\label{2.3}
\end{equation}
where coupling constants are defined as $J_{\perp} =K$ and $J_z = 2\Delta -K$, respectively, and $\vec{S}_i$ is a pseudospin operator of the localized electron in the orbital $i$ defined as
\begin{equation}
	\vec{S}_i = \frac{1}{2}\sum_{\sigma\sigma^{'}}f_{i\sigma}^{\dagger} \vec{\sigma}_{\sigma\sigma^{'}} f_{i\sigma^{'}}.
	\label{2.4}
\end{equation}
Furthermore, assuming that f-electrons constructing the f$^2$ state hybridize with conduction electrons which have the same symmetry as each $f^1$ state.
Thus the system can be described by the two-orbital impurity Anderson model with the ``antiferromagnetic Hund's-rule coupling'' as follows:
{\small
\begin{align}
	\label{2.5a}
	\mathcal{H} &= \mathcal{H}_{\rm c} + \mathcal{H}_{\rm hyb} + \mathcal{H}_{\rm f} + \mathcal{H}_{\rm Hund},\\
	\label{2.5b}
	\mathcal{H}_{\rm c} &=\sum_{i=1,2} \sum_{\vec{k}\sigma} \varepsilon_{\vec{k}} c_{\vec{k}i\sigma}^{\dagger} c_{\vec{k}i\sigma},\\
	\label{2.5c}
	\mathcal{H}_{\rm hyb} &= \sum_{i=1,2} \sum_{\vec{k}\sigma} \left( V_{i\vec{k}} c_{\vec{k}i\sigma}^{\dagger} f_{i\sigma} + {\rm h.c.} \right),\\
	\label{2.5d}
	\mathcal{H}_{\rm f} &=\sum_{i=1,2}\sum_{\sigma}E_{fi} f_{i\sigma}^{\dagger}f_{i\sigma} + \sum_{i=1,2}\sum_{\sigma}\frac{U_i}{2}f_{i\sigma}^{\dagger}f_{i \bar{\sigma}}^{\dagger} f_{i\bar{\sigma}}f_{i\sigma}, 
\end{align}
}
where $f_{i\sigma}(f_{i\sigma}^{\dagger})$ and $c_{\vec{k}i\sigma}(c_{\vec{k}i\sigma}^{\dagger})$ are annihilation (creation) operators of the f-electron on the orbital $i$ with the energy $E_{fi}$ and the conduction electron with wave vector $\vec{k}$ hybridizing with the f-electron with the symmetry of the orbital $i$ with strength $V_{i \vec{k}}$.
Here, the on-site intra-orbital Coulomb repulsion $U_i$ is explicitly taken into account, while other Coulomb repulsion terms like the inter-orbital or the exchange interaction, are implicitly included in the ``antiferromagnetic Hund's-rule coupling'' of (\ref{2.3}).\par
To analyze properties of the system described by the Hamiltonian (\ref{2.5a}) by the Wilson NRG method \cite{wilson,krishna}, we transform the conduction electron part as usual.
For simplicity, we take conduction bands to be isotropic in momentum space, i.e. the hybridization depends only on the orbital $i$, $V_{i\vec{k}} \equiv V_{i}$, and symmetric in the energy space (with an extent from $-D$ to $D$) about the Fermi level.
We discretize conduction bands logarithmically with the discretization parameter, $\Lambda$, and perform the unitary transformation assuming the density of state in conduction bands as constant.
Thus, eqs. (\ref{2.5b}) and (\ref{2.5c}) can be rewritten as
{\small
\begin{align}
	\mathcal{H}_{\rm c} &= \sum_{i,\sigma} \sum_{n=0}^{\infty} \Lambda^{-n/2} t_n \left( f_{i,n\sigma}^{\dagger} f_{i,n+1\sigma} +  f_{i,n+1\sigma}^{\dagger} f_{i,n\sigma}\right),\\
	\mathcal{H}_{\rm hyb} &= \sum_{i,\sigma} V_i \left( f_{i,0\sigma}^{\dagger} f_{i,-1\sigma} + f_{i,-1\sigma}^{\dagger} f_{i,0\sigma} \right),
	\label{2.6}
\end{align}
}
where $f_{i,n}$ ($f_{i,n}^{\dagger}$) is the annihilation (creation) operator of the conduction electron in the shell orbital whose extent is $k_{\rm F} \Lambda^{n/2}$ and $f_{i,-1\sigma} \equiv f_{i\sigma}$.
The hopping integral between $n$-th and $(n+1)$-th shell states, $t_n$, is expressed as
\begin{equation}
	t_n   = \frac{D(1+\Lambda^{-1})(1-\Lambda^{-n-1})}{2\sqrt{(1-\Lambda^{-2n-1})(1-\Lambda^{-2n-3})}}.
	\label{2.7}
\end{equation}
Then, we define $\mathcal{H}_N$ which approaches $\mathcal{H}/( D(1 + \Lambda^{-1})/2 )$ in the limit $N \rightarrow \infty$ as follows:
{\small
\begin{eqnarray}
\notag&&	\mathcal{H}_N = \Lambda^{(N-1)/2} \left[ \tilde{\mathcal{H}}_{\rm f} +  \sum_{i,\sigma} \tilde{V}_i\left( f_{i,0\sigma}^{\dagger} f_{i,-1\sigma} + f_{i,-1\sigma}^{\dagger} f_{i,0\sigma} \right) \right.\\
	\label{2.9}
	&& \left. + \sum_{i,\sigma} \sum_{n=0}^{N-1} \Lambda^{-n/2} \tilde{t}_n \left( f_{i,n\sigma}^{\dagger} f_{i,n+1\sigma} +  f_{i,n+1\sigma}^{\dagger} f_{i,n\sigma}\right) \right],
\end{eqnarray}
}
where the tilde indicates that energies are measured in a unit of $D(1+\Lambda^{-1})/2$.
The Hamiltonian (\ref{2.9}) satisfies the recursion relation 
\begin{equation}
	\mathcal{H}_{N+1} = \Lambda^{1/2}\mathcal{H}_N + \sum_{i\sigma} \tilde{t}_N \left(  f_{i,N\sigma}^{\dagger} f_{i,N+1\sigma} +  f_{i,N+1\sigma}^{\dagger} f_{i,N\sigma}\right).
	\label{2.10}
\end{equation}
We solve the whole sequence of Hamiltonian ($\mathcal{H}_{N}$) by using the recursive form (\ref{2.10}) with keeping states up to 1500 states in each iteration step, and use $\Lambda=3.0$ in all the calculations below unless explicitly stated. 
\section{NFL Behavior due to Competition between CEF and K-Y singlets}
In this section, we discuss the effect of the competition between the CEF singlet and the K-Y singlet states, which can give rise to a NFL state.
It is already known that the system described by the Hamiltonian (\ref{2.5a}) has the competition between the K-Y singlet and the $f^2$-CEF singlet states\cite{yotsuhashi}.
In general, the energy level and the strength of hybridization with conduction electron in each f-orbital are different.
In the present paper, we take parameters so that the Kondo temperature of orbital 2 is always lower than that of orbital 1: i.e., we set parameters of the two-orbital impurity Anderson model, eq (\ref{2.5a}), as $E_{f1}= E_{f2}=-0.4, U_1=U_2=1.0, V_1=0.45,$ and $V_2=0.3$.
Hereafter, the unit of energy is taken as $D$ unless stated explicitly.
In the case of $K=\Delta=0$, the model Hamiltonian, eq. (\ref{2.5a}), reduces to two independent impurity Anderson models.
The Kondo temperature of each orbital can be determined by the Wilson's definition, $4T_{\rm K}\chi_{\rm imp}(T=0) = 0.413$, for conventional Anderson model as $T_{\rm K1}= 6.10 \times 10^{-2}$ and $T_{\rm K2}= 6.01 \times 10^{-3}$, respectively.\par
\begin{figure}[t]
\begin{center}
	\includegraphics[width = 0.40\textwidth]{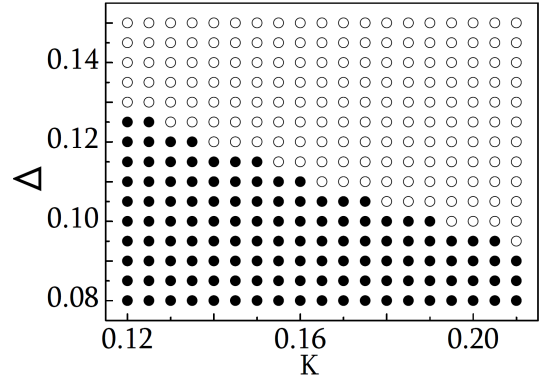}
	\caption{Phase diagram of the ground state in $K-\Delta$ plane. Filled circles represent the K-Y singlet fixed point and open circles represent the $f^2$-CEF singlet fixed point. Parameter set is $E_{f1}=E_{f2}=-0.4, U_1=U_2=1.0, V_1=0.45,$ and $V_2=0.3$. }
	\label{fig2} 
\end{center}
\vspace{-8mm}
\end{figure}

For the finite value of CEF parameters, $(K,\Delta)$, there are two stable Fermi Liquid (FL) fixed points corresponding to two singlet ground states as shown in Fig.\ref{fig2}: the K-Y singlet (filled circles) and the CEF singlet (open circles) fixed points.
At the boundary of these two regions of FL fixed points, there exists a curve of critical points, across which energy spectra for even and odd iteration interchange, and NFL behaviors appear in the vicinity of the boundary.
To analyze further, we fix one of the CEF parameters as $K=0.16$, and calculate the physical properties for a series of values of the CEF splitting parameter $\Delta$.
Analyzing near the critical point in more detail, the critical value of $\Delta$ is determined as $\Delta^{*} \simeq 0.112$ for $K=0.16$.\par
\begin{figure}[t]
\begin{center}
	\includegraphics[width = 0.40\textwidth]{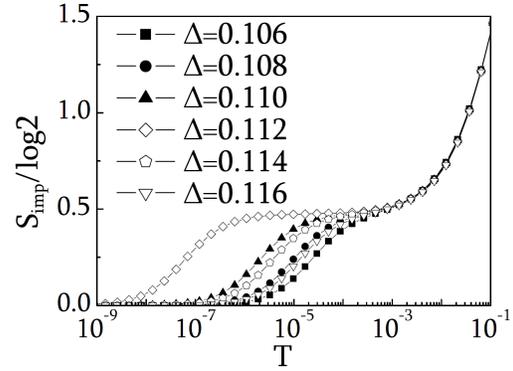}
	\caption{Temperature dependence of the entropy due to f-electrons in systems near the critical point. Parameter set is the same as that used in Fig.\ref{fig2}. In order to obtain the result with a higher accuracy, 3000 states are kept in each step of NRG. Ground states of each system are indicated by open symbols for the K-Y singlet, and filled symbols for the CEF singlet. The characteristic temperature $T_{\rm F}^{*}$ is given by that making $\partial S_{\rm imp}(T)/\partial ( \log\,T)$ maximum at the lower temperature side.}
	\label{fig3} 
\end{center}
\vspace{-8mm}
\end{figure}

Fig.\ref{fig3} shows the result of the temperature dependence of $S_{\rm imp}(T)$, the entropy due to f-electrons, near the critical point. 
As mentioned above, the characteristic temperature $T_{\rm F}^{*}$ is defined as the temperature at which the temperature derivative of entropy, $\partial S_{\rm imp}(T)/\partial (\log\,T)$, takes the maximum value just before $S_{\rm imp}(T)$ approaching 0 as $T \rightarrow 0$.
As seen in Fig.\ref{fig3}, $T_{\rm F}^{*}$ is drastically suppressed by the effect of the competition near the critical value of CEF splitting $\Delta =0.112 \simeq \Delta^{*}$.\par
\begin{figure}[tb]
\begin{center}
	\includegraphics[width = 0.38\textwidth]{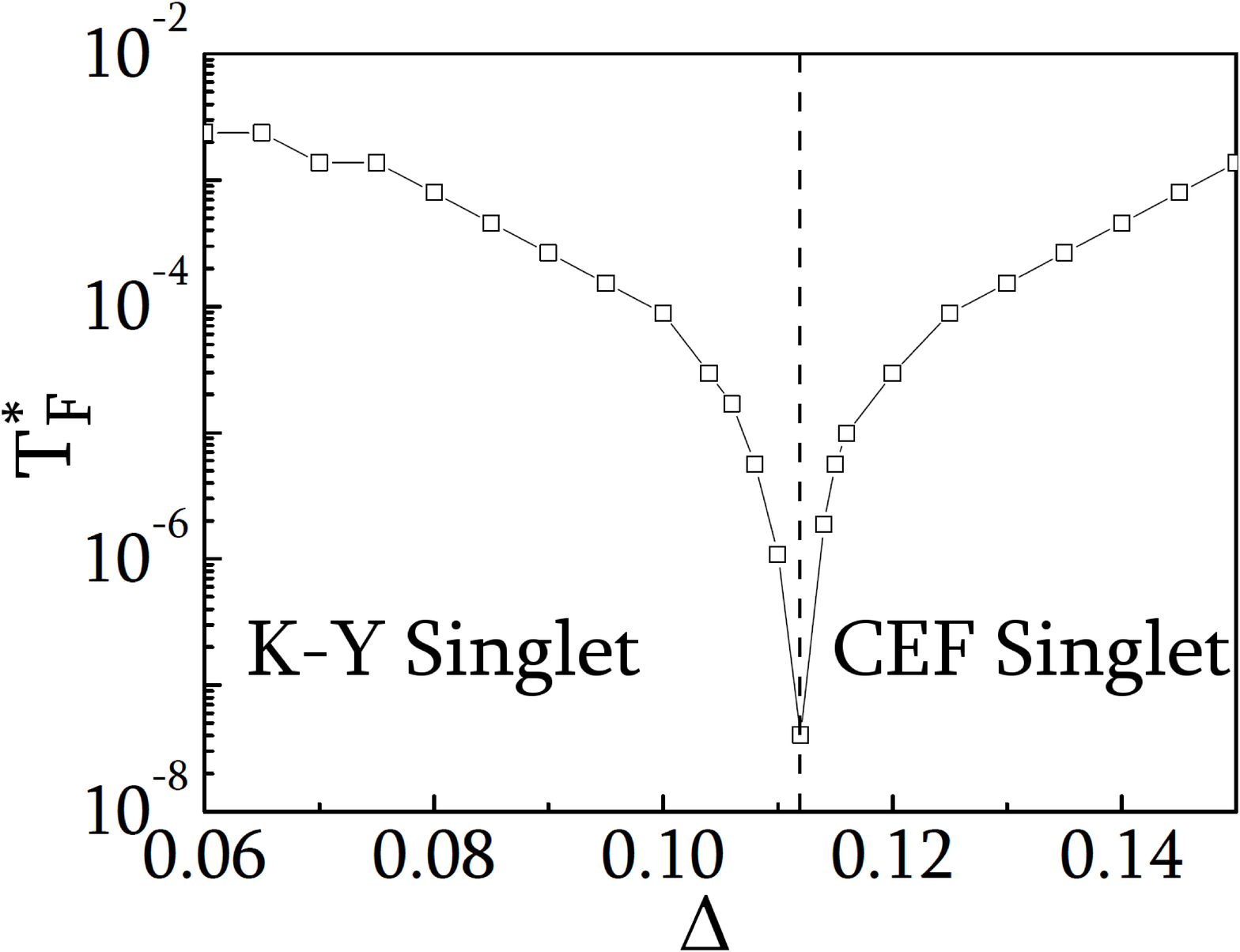}
	\caption{$\Delta$ dependence of $T_{\rm F}^{*}$. The effect of the competition between two singlet states suppresses $T_{\rm F}^{*}$, and in particular $T_{\rm F}^{*} =0$ at the critical point $\Delta = 0.112 \simeq \Delta^{*}$. }
	\label{fig4} 
\end{center}
\vspace{-6mm}
\end{figure}

Fig.\ref{fig4} shows the $\Delta$ dependence of $T_{\rm F}^{*}$ which is obtained by numerical calculations of $S_{\rm imp}(T)$. 
In the case of $\Delta < \Delta^{*}$, the K-Y singlet state is the ground state, and two localized moments $\vec{S}_1$ and $\vec{S}_2$ are screened out independently by corresponding conduction electrons, where each Kondo temperature is affected by the interaction between f-electrons.
In this case, the total phase shift of conduction electrons characterizing this fixed point is $\delta=\pi\ (\delta_1=\pi/2, \delta_2 =\pi/2)$, and $T_{\rm F}^{*}$ is given by a value slightly lower than the Kondo temperature $T_{\rm K2}$, if $\Delta$ is much smaller than $\Delta^{*}$.
On the other hand, in the case of $\Delta >\Delta^{*}$, the CEF splitting (antiferromagnetic interaction between f-electrons in the model Hamiltonian, (\ref{2.5a})) is so large compared to the energy gain related to the formation of K-Y singlet states that the CEF singlet becomes the ground state.
In this case, the remaining conduction electrons are not scattered by f-electrons, and as a result the total phase shift is $\delta=0\ (\delta_1=0,\delta_2=0)$.
When $\Delta \gg \Delta^{*}$, $T_{\rm F}^{*}$ becomes close to the excitation energy $K$ between two singlet states.\par
\begin{figure}[t]
\begin{center}
	\includegraphics[width = 0.30\textwidth]{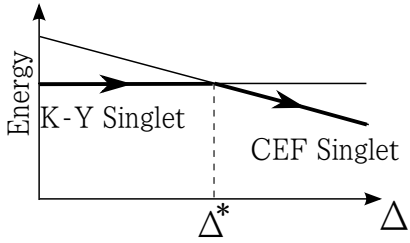}
	\caption{Schematic energy levels of two singlet ground states. The CEF singlet state is stabilized relative to the K-Y singlet state as $\Delta$ increases. }
	\label{fig5} 
\end{center}
\vspace{-7mm}
\end{figure}

Such an interchange of the ground state can be understood by considering that the increase of $\Delta$ causes the stabilization of the level of the CEF singlet state as shown in Fig.\ref{fig5}.
In the case of $\Delta \sim \Delta^{*}$, $T_{\rm F}^{*}$ is determined not by characteristic energies of the K-Y singlet and the CEF singlet states, but by the energy splitting between two singlet states, $\Delta E$: i.e., $T_{\rm F}^{*} \sim \Delta E$. 
Particularly, at the critical point, the degeneracy of the K-Y singlet and the CEF singlet states is not lifted even at $T =0$, making $T_{\rm F}^{*}=0$ and $\lim_{T \rightarrow 0}S_{\rm imp} = 0.5 \log\,2$.
In other words, at low enough temperatures, the localized moment $\vec{S}_1$ of orbital 1 has already been screened out by conduction electrons in orbital 1 below $T_{\rm K1}$, while $\vec{S}_2$ of orbital 2 still has the degree of freedom as localized moment.
Therefore, the effective Hamiltonian of (\ref{2.5a}) near the fixed point behaves as the two-channel Kondo model (TCKM) \cite{cragg,pang} because $\vec{S}_2$ interacts with two ``conduction'' electron channels, one is the conduction electrons on orbital 2 and the other is a complex of conduction electrons on orbital 1 and screened $\vec{S}_{1}$ as discussed in ref. 10.\par
\begin{figure}[t]
\begin{center}
	\includegraphics[width = 0.4\textwidth]{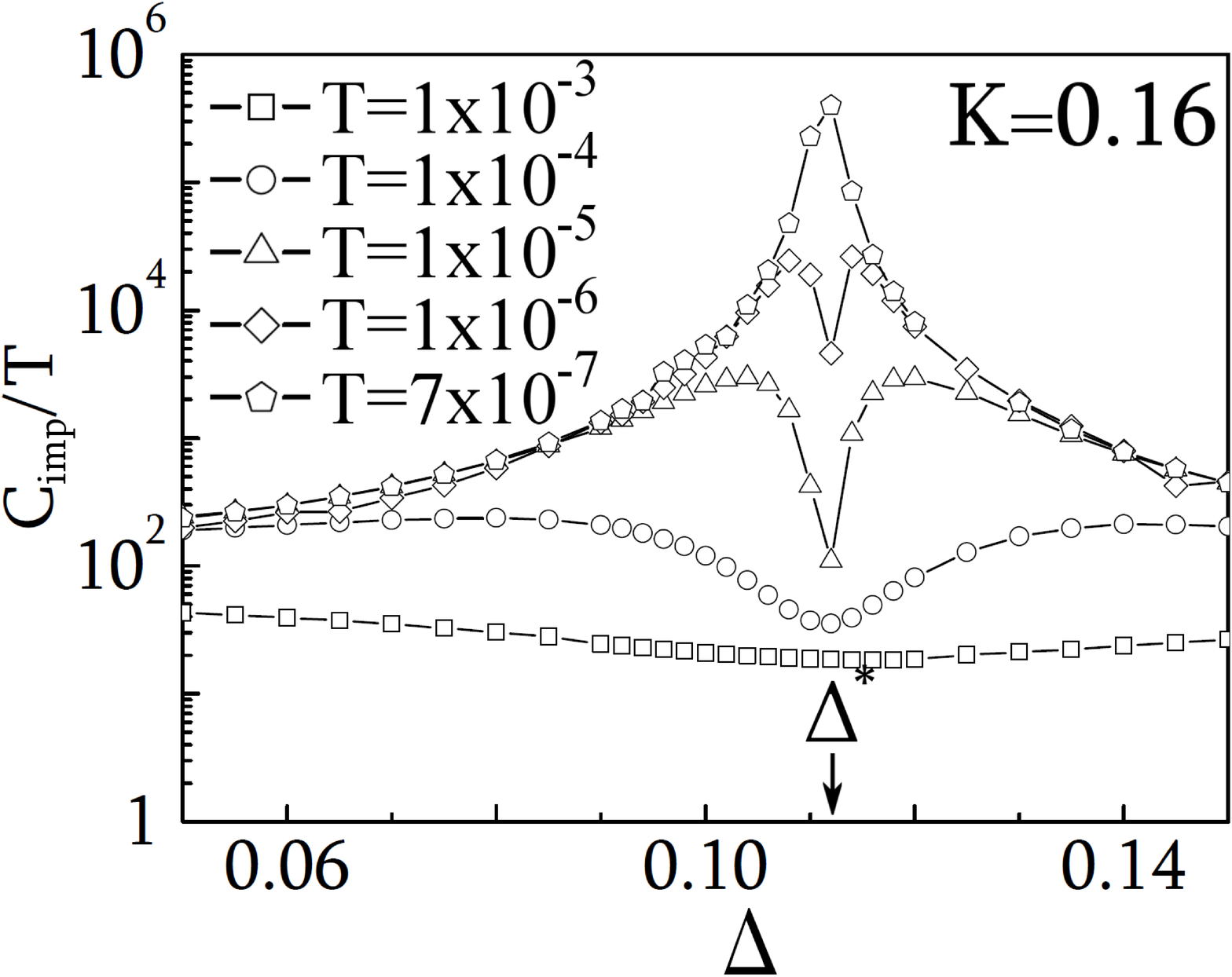}
	\caption{$\Delta$ dependence of the Sommerfeld coefficient $\gamma_{\rm imp} (T) \equiv C_{\rm imp}(T)/T$ due to the f-electrons for various temperature. The ground state switch at $\Delta =\Delta^{*} \simeq 0.112$ from the K-Y singlet ground state for $\Delta < \Delta^{*}$ to the CEF singlet for $\Delta > \Delta^{*}$. }
	\label{fig6} 
\end{center}
\vspace{-7mm}
\end{figure}

Fig.\ref{fig6} shows the $\Delta$ dependence of the Sommerfeld coefficient, $\gamma_{\rm imp}(T) \equiv C_{\rm imp}(T)/T$, due to f-electrons for various temperatures.
For all $\Delta$ shown in Fig.\ref{fig6}, $\gamma_{\rm imp}(T)$ increases monotonically down to $T=7.0\times 10^{-7}$ as decreasing $T$.
At $\Delta = \Delta^{*} \simeq 0.112$, the increase of $\gamma_{\rm imp}(T)$ does not stop and exhibits divergence in the limit $T \rightarrow 0$ because the structure of the fixed point is the same as that of TCKM as discussed above.
For $\Delta$ off the critical value $\Delta^{*}$, the increase of $\gamma_{\rm imp}(T)$ stops at around the characteristic temperature $T_{\rm F}^{*}$ leading to the Fermi liquid behavior at $T < T_{\rm F}^{*}$.
$\gamma_{\rm imp}(T)$ takes a dip structure around $\Delta \sim \Delta^{*}$ at higher temperature region.
This is because $S_{\rm imp}(T)$ has only a weak $T$ dependence in a wide temperature range $0 \sim T_{\rm F}^{*} < T < T_{\rm K2}$ or $\Delta$ around $\Delta \simeq \Delta^{*}$ as can be seen in Fig.\ref{fig3}.\par
It is remarked that the enhanced part of $\gamma_{\rm imp}(T)$ near $\Delta \sim \Delta^{*}$ in the low temperature limit from the background part at $\vert \Delta - \Delta^{*} \vert \gg  \Delta^{*}$ arises from the effect of the competition between the K-Y singlet and the CEF singlet states.
The part of the background is essentially given by an inverse of $T_{\rm K2}$ or $\Delta$, and is overwhelmed by the enhanced part near $\Delta \sim \Delta^{*}$.
Note that the ordinate of Fig.\ref{fig6} is represented in a logarithmic scale.\par
Although we take $\Delta$ as a control parameter here, we can expect a similar behavior of $\gamma_{\rm imp}$ through other parameters, such as the hybridizations $V_1$ and $V_2$, which can also control the competition between levels of two singlet states.
It is also remarked that such an anomalous behavior of $\gamma_{\rm imp}$ can be realized in systems with other symmetry: e.g., in UBe$_{13}$ with cubic symmetry\cite{ott,ott2, gegenwart}.
In this material, $\gamma$ shows the similar behavior as shown in Fig.\ref{fig6} through the change of the lattice constant, $a_{0}$, which is controlled by replacing the U atom partly with other nonmagnetic elements.
It is remarkable that $\gamma$ takes a maximum value at $a_{0} = a_{0}^{*}$, which is approximately the same as the lattice constant of UBe$_{13}$\cite{kim}.
Experimentally, in a series of materials with $a_{0} < a_{0}^{*}$, the Kondo like upturn is observed in the resistivity in the low temperature region, while in those with $a_{0} > a_{0}^{*}$, the temperature dependence of the resistivity can be explained by the effect of the CEF with the singlet ground state.
Then, we expect that UBe$_{13}$ is located near the critical point in this series of materials.
\section{Magnetic Field Dependence of Non-Fermi Liquid Behavior}
In this section, we discuss the magnetic field dependence of the NFL behavior of $\gamma_{\rm imp}(T)$.
The effect of the magnetic field on $f^1$ states is taken into account through the Zeeman term for total angular moment, $\mathcal{H}_{\rm Zeeman}(f^1) = - g_J \mu_{\rm B} j_z H_z$, with $j=5/2$ and $g_j=6/7$.
That on $f^2$ states arises from the diagonal (for $\Gamma_5^{(2)}$ doublet) and the off-diagonal (for $\Gamma_3$ and $\Gamma_4$ singlets) matrix elements of two-electron Zeeman term $\mathcal{H}_{\rm Zeeman} (f^2)$; e.g., $\langle \Gamma_{5 \pm }^{(2)} \vert \mathcal{H}_{\rm Zeeman} (f^2) \vert \Gamma_{5 \pm}^{(2)} \rangle=\mp 11 g_j \mu_{\rm B} H_z /7$, and $\langle \Gamma_3 \vert \mathcal{H}_{\rm Zeeman} (f^2) \vert \Gamma_4 \rangle = -2 g_j \mu_{\rm B} H_z$.
Here, $\mathcal{H}_{\rm Zeeman}(f^2)$ consists of two $\mathcal{H}_{\rm Zeeman}(f^1)$.\par
\begin{figure}[t]
\begin{center}
	\includegraphics[width = 0.45\textwidth]{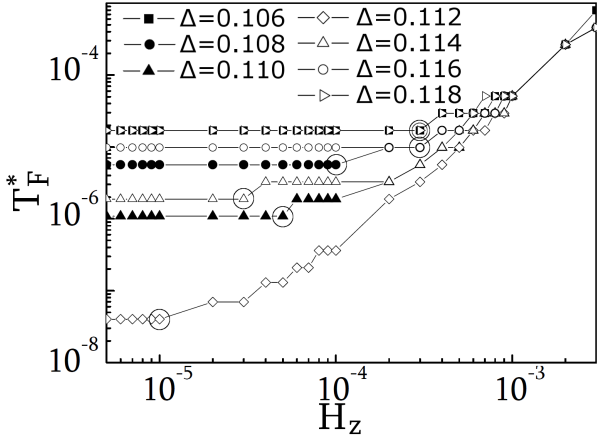}
	\caption{Magnetic field dependence of $T_{\rm F}^{*}$ near the critical point. The parameters related to f-electrons are the same as Fig.\ref{fig2}. Circles indicate characteristic magnetic fields $Hz^{*}$'s.}
	\label{fig7} 
\end{center}
\vspace{-7mm}
\end{figure}

In Fig.\ref{fig7}, we show the magnetic field dependence of the characteristic temperature $T_{\rm F}^{*}$ near the critical point; i.e., $\Delta=0.108, 0.110, 0.112(\simeq \Delta^{*}), 0.114, 0.116$ and $0.118$.
It is noted that $T_{\rm F}^{*}(H_z)$ remains constant for $H_z$ less than the characteristic magnetic field $H_z^{*}$ which is defined approximately as that from which $T_{\rm F}^{*}( H_z)$ starts to increase as increasing $H_z$ (as shown by circles in Fig.\ref{fig7}).
Explicitly, the characteristic magnetic field $H_{z}^{*}$'s are given as $H_z^{*} \simeq 3 \times 10^{-4}$ for $\Delta = 0.106$ and $0.118$, $H_z^{*} \simeq 2 \times 10^{-4}$ for $\Delta = 0.108$ and $0.116$, $H_{z}^{*} \simeq 3 \times 10^{-5}$ for $\Delta = 0.110$ and $0.114$, and $H_z^{*} \simeq 1 \times 10^{-5}$ for $\Delta=0.112$.
$H_z^{*}$ has a tendency of approaching zero as the critical fixed point is approached, i.e., $\Delta \rightarrow \Delta^{*}$.
For CEF parameter $\Delta$ shown in Fig.\ref{fig7}, $H_z^{*}$ is much smaller than the lower Kondo temperature $T_{\rm K2} \simeq 6.01 \times 10^{-3}$, so that the magnetic field $H_z < H_{z}^{*}$ has little influence on the K-Y singlet state.
Then, $H_{z}^{*}$ is considered to be determined by a competition of two effects which destroy the TCKM-type NFL fixed point: one is a distance of $\Delta$ from $\Delta^{*}$ and the other is the magnetic field which breaks the degeneracy corresponding to $S_{\rm imp}(T=0) = 0.5 \log\,2$ due to the TCKE, the origin of the TCKM-type NFL fixed point.
Namely, $H_z^{*}$ is given by the energy scale characterizing a crossover from the TCKM-type NFL behavior to the polarized Fermi liquid behavior beyond the effect of the distance of $\Delta$ from the critical value $\Delta^{*}$.
Since $\gamma_{\rm imp}(T)$ exhibits the divergent increase around $\Delta \sim \Delta^{*}$ in the temperature region $T > T_{\rm F}^{*}(H_z)$ as decreasing $T$, $\gamma_{\rm imp}(T)$ exhibits a NFL behavior in the same temperature region $T > T_{\rm F}^{*}(H_z)$.
Since $T_{\rm F}^{*}(H_z)$ remains almost unchanged up to $H_z = H_z^{*}$, the NFL behaviors are expected to remain robustly even under the magnetic field $H_z > T_{\rm F}^{*}(H_z)$ so long as $H_z < T_{\rm K2}$.
This behavior is reproduced by explicit calculations of $\gamma_{\rm imp}(T)$ under various magnetic fields as shown below.\par
In Fig.\ref{fig8}, we show the temperature dependence of $\gamma_{\rm imp}(T)$ for $\Delta= 0.112$ $(\simeq \Delta^{*})$ and $\Delta=0.118$ under various magnetic fields of up to $H_z = 1.2 \times 10^{-3}$. 
\begin{figure}[t]
\begin{center}
	\includegraphics[width = 0.4\textwidth]{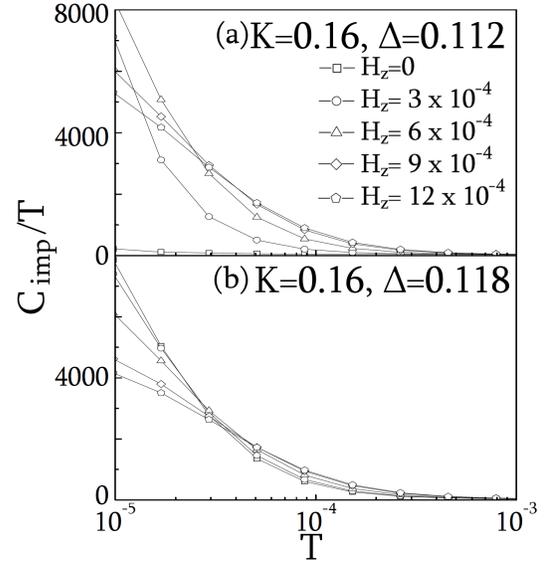}
	\caption{Temperature dependence of $\gamma$ for (a) $\Delta= 0.112$ ($\simeq \Delta^{*}$) and (b) $\Delta=0.118$ under various magnetic fields. In the case of (b), the NFL behavior of $\gamma$ is robust against a magnetic field of up to $H_z = 1.2 \times 10^{-3}$ for $T>3.0 \times 10^{-5}$ in spite of $T_{\rm F}^{*} \simeq 1.69 \times 10^{-5}$. The parameters related to f-electrons are the same as those used in Fig.\ref{fig2}. }
	\label{fig8} 
\end{center}
\vspace{-7mm}
\end{figure}
Extremely close to the criticality at $\Delta = 0.112 \simeq \Delta^{*}$, $\gamma_{\rm imp}(T)$ is enhanced by the magnetic field as shown in Fig.\ref{fig8}(a).
This is because $T_{\rm F}^{*}(H_z)$ increases appreciably from $10^{-7}$ to $10^{-5}$ corresponding to the increase of the magnetic field $H_z$ from $10^{-4}$ to $10^{-3}$, resulting in an increase of $\partial S_{\rm imp}(T)/ \partial (\log\,T) = C_{\rm imp}(T)$, so $\gamma_{\rm imp}(T)$, at $T > 10^{-5}$.
On the other hand, at $\Delta = 0.118$ slightly off the criticality, $\gamma_{\rm imp}(T)$ is robust against the magnetic field up to $H_z =1.2 \times 10^{-3}$ for the temperature region $T>3 \times 10^{-5}$ as shown in Fig.\ref{fig8}(b).
This is because $T_{\rm F}^{*}(H_z)$ remains almost unchanged up to $H_z = H_z^{*} \sim 10^{-3}$ so that $\gamma_{\rm imp}(T)$ remains the same as that at $H_z = 0$ for $T > T_{\rm F}^{*} \simeq 10^{-5}$.\par
These kinds of NFL behaviors arise also in the region of the K-Y singlet state, i.e., $\Delta < \Delta^{*}$, although we do not show the results explicitly.
\section{Kondo-Temperature Dependence of Non-Fermi Liquid Behavior under Magnetic Field}
\begin{figure}[t]
\begin{center}
	\includegraphics[width = 0.4\textwidth]{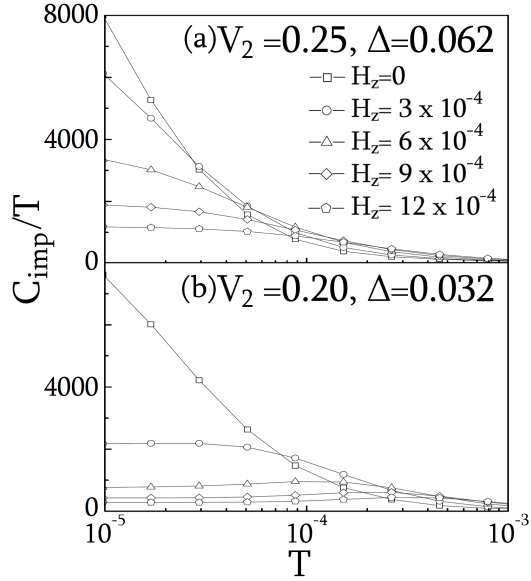}
	\caption{Temperature dependence of $\gamma_{\rm imp}(T)$ for a series of magnetic fields in the system for the hybridization (a)$V_2=0.25$ and (b)$V_2=0.20$. }
	\label{fig9}
\end{center}
\vspace{-6mm}
\end{figure}

In this section, we investigate the properties of the NFL behavior of $\gamma_{\rm imp}(T)$ under magnetic field of systems with other $T_{\rm K2}$ by changing $V_2$ as $V_2=0.25$ and $0.20$ for various sets of the CEF parameter, $(K,\Delta)$.
Other parameters are set to be the same as those in the previous section: i.e., $E_{f1}= E_{f2} =-0.4, U_1=U_2=1.0$, and $V_1=0.45$.
In the case of $K=\Delta=0$, each lower Kondo temperature can also be determined by the Wilson's definition as $T_{ {\rm K2}}=1.27 \times 10^{-3}$ for $V_2=0.25$ and $T_{\rm K2}=8.92 \times 10^{-5}$ for $V_2=0.20$, respectively. 
To analyze further, we also fix one of the CEF parameters as $K=0.16$ and calculate $\gamma_{\rm imp}(T)$ for a series of $\Delta$ under various magnetic fields. 
It is natural that $\Delta^{*}$ (corresponding to the critical point) becomes small with decreasing $T_{\rm K2}$ because the energy gain due to the formation of the K-Y singlet state decreases with a smaller $V_2$.
The critical value of $\Delta$ is determined as $\Delta^{*} \simeq 0.054$ for $V_2=0.25$ and $\Delta^{*} \simeq 0.024$ for $V_2=0.20$, respectively.
Fig.\ref{fig9} shows the temperature dependence of $\gamma_{\rm imp}(T)$ of the system with the CEF ground state: (a) $\Delta = 0.062 > \Delta^{*} \simeq 0.054$ for $V_2 = 0.25$ and (b) $\Delta = 0.032 > \Delta^{*} \simeq 0.024$ for $V_2=0.20$.
The NFL behavior being robust against the magnetic field occurs in a temperature region of $T> T_{\rm F}^{*}$ up to $H_z \simeq H_z^{*}$ in the former case (a), while in the latter case (b) the magnetic field has considerable influence on the NFL behavior.\par
For $H_z =0$, $T_{\rm F}^{*}$ is also suppressed as in the case of $V_2=0.30$ in the vicinity of the critical point $\Delta \sim \Delta^{*}$.
It is noted that the decrease of $T_{\rm K2}$ and $\Delta$ does not appreciably affect $T_{\rm F}^{*}$, i.e. $T_{\rm F}^{*} \sim 10^{-5}$ for both cases of (a) and (b), which is determined from calculations corresponding to Fig.\ref{fig3}.
This is because $T_{\rm F}^{*}$ is determined by the energy splitting between the K-Y singlet and the CEF singlet states, and does not depend on the characteristic energy scale of each singlet state.  
Under the magnetic field, the effect on the NFL behavior is markedly different in two cases (a) and (b).
In the case of (a) with $V_2 = 0.25$, the NFL behavior of $\gamma_{\rm imp}(T)$ is rather robust against the magnetic field (up to $H_z^{*}$) in a wide temperature range ($T>T_{\rm F}^{*}$) as in the case of $V_2 = 0.30$, while in the case of (b) with $V_2=0.20$, $\gamma_{\rm imp}(T)$ is sensitive to the magnetic field because the characteristic magnetic field $H_z^{*}$ is comparable to the lower Kondo temperature, $T_{\rm K2}$.
Namely, in the case of $V_2=0.20$, the magnetic field $H_z > T_{\rm K2} \simeq 8.92 \times 10^{-5}$ suppresses $\gamma_{\rm imp}(T)$ by breaking the K-Y singlet ground state.
It is noted that in the case $T_{\rm K2} > \Delta $, the suppression of $\gamma_{\rm imp}(T)$ as in the case of Fig.\ref{fig8}(b) is expected for $H_z > \Delta$ by breaking the CEF singlet states.
Thus, the magnetic field dependence of the NFL behavior of $\gamma_{\rm imp}(T)$ is determined not by the characteristic temperature $T_{\rm F}^{*}$, but by the characteristic magnetic field $H_z^{*}$ which is determined by the characteristic energy scale of each singlet state, $T_{\rm K2}$ and $\Delta$, or the distance from the critical point.
\section{Conclusion and Discussion}
We have investigated the effect of the magnetic field on the NFL behaviors due to the competition between the K-Y singlet and the CEF singlet states in f$^2$-based heavy fermion systems with tetragonal symmetry.
The effect of the competition suppresses the characteristic temperature $T_{\rm F}^{*}$, corresponding to a peak of the specific heat, $C_{\rm imp}(T)$, to a much smaller value than the characteristic energy scale of each singlet states: i.e., $T_{\rm K2}$, the lower Kondo temperature, and $\Delta$, the energy splitting between the CEF singlet ground state and the first excited doublet states.
$T_{\rm F}^{*}$ is determined approximately by $\Delta E$, the energy difference between two singlet states, and there exists the two-channel Kondo model (TCKM) type NFL behaviors at $T_{\rm F}^{*} < T < T_{\rm K2}$.
Namely, near the critical point, $\Delta \sim \Delta^{*}$, the Sommerfeld coefficient $\gamma_{\rm imp}(T)$ exhibits a NFL behavior ($\gamma_{\rm imp}(T) \propto -\log\,T$) at $T > T_{\rm F}^{*}$.\par
In the vicinity of the critical point, $T_{\rm F}^{*}$ was shown not to be affected by the magnetic field up to a certain value $H_z^{*}$, while $T_{\rm F}^{*}$ is increased for $H_z^{*} < H_z < {\rm min}(T_{\rm K2}, \Delta)$.
As a result, the NFL behavior of $\gamma_{\rm imp}$ at $T>T_{\rm F}^{*}$ is robust against the magnetic field $H<H_{z}^{*}$.
Then, for reasonable sets of parameters, the NFL behaviors being robust against a magnetic field of up to $H_{z}^{*} $ can occur at an observable temperature range.
Thus, the magnetic field dependence of this NFL is characterized by $H_z^{*}$ which is determined by the characteristic energy scales of two singlet states and the distance from the critical point.\par
In the present paper, we have discussed physical properties in the tetragonal symmetry.
However, also in the case of other crystal symmetries, it is expected that there remains the effect of the competition between the K-Y singlet and the CEF singlet states, leading to the NFL behaviors similar to the present case.
One example would be the case of the cubic system UBe$_{13}$ which seems to be located near the phase boundary between the K-Y singlet and the CEF singlet states, according to a series of experiments of $\lim_{T \sim 0} C(T)/T $ for systems of solid solution, U$_{1-x}$T$_{x}$Be$_{13}$, where the lattice constant $a_0$ is changed in a wide range covering both the K-Y singlet and the CEF singlet ground states\cite{kim}.
Moreover, pure UBe$_{13}$ exhibits the NFL behavior, $C(T)/T \sim -\log\,T$ up to $H_z = 12$ Tesla\cite{gegenwart}.
Of course, precisely speaking, results of the present paper are for the system of $f^2$-impurity so that we should be careful in deriving a solid conclusion.
Indeed, an approach based on the dynamical mean field concept is indispensable for deriving a solid conclusion for lattice systems, in which the present results would be inherited to the solver of impurity problem.
Nevertheless, we expect that the effect of the competition plays an important role for UBe$_{13}$ to exhibit such a NFL behavior rather robust against the magnetic field larger than the effective Fermi energy inferred from the value of $\lim_{T \sim 0}C(T)/T$.
Namely, the lower Kondo temperature would be larger than 12K from the fact that the NFL behavior $\lim_{T\sim 0}C(T)/T \sim - \log\,T$ in UBe$_{13}$ is robust against the magnetic field up to 12 Tesla at least\cite{gegenwart}.
Were it not for the superconducting state at $T<T_{\rm c} \simeq 1$K, there would exist the peak with specific heat near at $T = T^{*}_{\rm F}$.
Predictions of the present paper may be checked by experiments in some U-diluted system of UBe$_{13}$ near the phase boundary between the K-Y singlet and the CEF singlet states under pressures and/or magnetic fields.
\section*{Acknowledgements}
We are grateful to K. Hattori and S. Yotsuhashi for stimulating conversations and discussions.
One of the authors (K.M.) is grateful to T. Kasuya for directing his attention to ref.15 on an occasion of the workshop of a Grant-in-Aid for Scientific Research on Priority Areas ``Filled Skutterudites'' held at Tokyo Metropolitan University in November, 2003.
S.N. and H.M. are supported by the Global COE program (G10) from The Japan Society for the Promotion of Science.
This work is supported by a Grant-in-Aid for Scientific Research on Innovative Area ``Heavy Electrons'' (No.20102008) from the Ministry of Education, Culture, Sports, Science and Technology. 

\appendix
\section{$f^2$ States in Tetragonal Symmetry}
In the tetragonal symmetry, wave functions for each CEF level are given within f$^1$ states of $j=5/2$ orbitals as follows:
{\small
\begin{eqnarray}
	\label{gamma71}
	\vert \Gamma_{7,\pm}^{(1)} \rangle &=& \zeta \vert \pm \frac{5}{2} \rangle  + \eta \vert \mp \frac{3}{2} \rangle, \\
	\label{gamma72}
	\vert \Gamma_{7,\pm}^{(2)} \rangle &=& \pm \eta \vert \pm \frac{5}{2} \rangle  \mp \zeta \vert \mp \frac{3}{2} \rangle, \\
	\label{gamma6}
	\vert \Gamma_{6,\pm} \rangle &=& \vert \pm \frac{1}{2} \rangle,
\end{eqnarray}
}
where $\zeta$ and $\eta$ are the coefficients determined by the effect of the CEF.
In this appendix, we determine these coefficients on the basis of the condition that the energy level of low-lying f$^2$ states with $J=4$ manifold can be reproduced by Hamiltonian (\ref{2.3}).
First, we construct states with $J=4$ manifold from the direct product of states with $j=5/2$ manifold.
{\small
\begin{align}
	\vert \pm 3 \rangle &= \pm \vert \pm \frac{5}{2} \rangle \vert \pm \frac{1}{2} \rangle, \\
	\vert \pm 2 \rangle  &= \pm \frac{3}{\sqrt{14}} \vert \pm \frac{5}{2} \rangle \vert \mp \frac{1}{2} \rangle \pm \sqrt{\frac{5}{14}} \vert \pm \frac{3}{2} \rangle \vert \pm \frac{1}{2} \rangle, \\
	\vert \pm 1 \rangle &= \pm \sqrt{\frac{2}{7}} \vert \pm \frac{5}{2} \rangle \vert \mp \frac{3}{2} \rangle \pm \sqrt{\frac{5}{7}} \vert \pm \frac{1}{2} \rangle \vert \mp \frac{3}{2} \rangle.
	\label{append3}
\end{align}
}
By using the inversion relation of eqs.(\ref{gamma71})-(\ref{gamma6}) representing $\vert \pm 5/2 \rangle$, $\vert \pm 3/2 \rangle$, and $\vert \pm 1/2 \rangle$ in terms of $\Gamma_{7}^{(1)}$, $\Gamma_{7}^{(2)}$, and $\Gamma_{6}$, we obtain the f$^2$ states in the tetragonal symmetry as follows:
{\small
\begin{align}
\notag	\vert \Gamma_4 \rangle &= \frac{1}{\sqrt{2}} \left( \vert 2 \rangle - \vert -2 \rangle \right)\\
\notag &= \frac{1}{2 \sqrt{7}} \left[ \left( 3 \zeta + \sqrt{5} \eta \right) \left( \vert \Gamma_{7+}^{(1)} \rangle \vert \Gamma_{6-} \rangle  +  \vert \Gamma_{7-}^{(1)} \rangle \vert \Gamma_{6+} \rangle \right) \right.\\
	\label{gamma4}
	& \left. + \left(  \sqrt{5} \zeta - 3 \eta \right) \left( \vert \Gamma_{7-}^{(2)} \rangle \vert \Gamma_{6+} \rangle - \vert \Gamma_{7+}^{(2)} \rangle \vert \Gamma_{6-} \rangle \right) \right],\\
\notag	\vert \Gamma_3 \rangle &= \frac{1}{\sqrt{2}} \left( \vert 2 \rangle + \vert -2 \rangle \right)\\
\notag &= \frac{1}{2 \sqrt{7}} \left[ \left( 3 \zeta - \sqrt{5} \eta \right) \left( \vert \Gamma_{7+}^{(1)} \rangle \vert \Gamma_{6-} \rangle  -  \vert \Gamma_{7-}^{(1)} \rangle \vert \Gamma_{6+} \rangle \right) \right.\\
	\label{gamma3}
& \left. + \left(  \sqrt{5} \zeta + 3 \eta \right) \left( \vert \Gamma_{7+}^{(2)} \rangle \vert \Gamma_{6-} \rangle + \vert \Gamma_{7-}^{(2)} \rangle \vert \Gamma_{6+} \rangle \right) \right],\\
\notag	\vert \Gamma_{5,+}^{(2)} \rangle &= \beta \vert 3 \rangle - \alpha \vert -1 \rangle\\
\notag &= \left( \beta \zeta + \sqrt{\frac{5}{7}} \alpha \eta \right) \vert \Gamma_{7+}^{(1)}\rangle \vert \Gamma_{6+} \rangle + \sqrt{\frac{2}{7}} \alpha \vert \Gamma_{7-}^{(1)} \rangle \vert \Gamma_{7 - }^{(2)} \rangle \\
	\label{gamma5+}
	& + \left( \beta \eta - \sqrt{\frac{5}{7}} \alpha \zeta \right) \vert \Gamma_{7+}^{(2)} \rangle  \vert \Gamma_{6+} \rangle, \\ 
\notag	\vert \Gamma_{5,-}^{(2)} \rangle &= \beta \vert - 3 \rangle - \alpha \vert 1 \rangle \\
\notag &=  - \left( \beta \zeta + \sqrt{\frac{5}{7}} \alpha \eta \right) \vert \Gamma_{7-}^{(1)}\rangle \vert \Gamma_{6-} \rangle + \sqrt{\frac{2}{7}} \alpha \vert \Gamma_{7+}^{(1)} \rangle \vert \Gamma_{7 + }^{(2)} \rangle \\
	\label{gamma5-}
	& + \left( \beta \eta - \sqrt{\frac{5}{7}} \alpha \zeta \right) \vert \Gamma_{7-}^{(2)} \rangle  \vert \Gamma_{6-} \rangle,  
\end{align}
}
are the same as those expressed in eqs. (\ref{2.2a})-(\ref{2.2d}).\par
Here, terms where states with $\Gamma_{7}^{(1)}$ symmetry are occupied in eqs.(\ref{gamma4})-(\ref{gamma5-}) can be negligible because their energy levels are assumed to be higher than the other states so that the hybridization between $\Gamma_{7}^{(1)}$ and the f$^2$-states (\ref{gamma4})-(\ref{gamma5-}) may be neglected for forming a heavy fermion state as discussed in refs. 16 and 17.
The coefficients, $\zeta$ and $\eta$, can be determined by the condition that the coefficients of the remaining terms in eqs.(\ref{gamma4})-(\ref{gamma3}) are equal to those in eqs.(\ref{2.2a})-(\ref{2.2b}).
The result is 
\begin{equation}
	\zeta = \sqrt{\frac{5}{14}},
	\hspace{15mm}\eta = \frac{3}{\sqrt{14}}.
	\label{append4}
\end{equation}
A relation between $\alpha$ and $\beta$ is also derived by comparing eqs.(\ref{gamma5+})-(\ref{gamma5-}) with eqs.(\ref{2.2c})-(\ref{2.2d}) as follows:
\begin{equation}
	\beta \frac{3}{\sqrt{14}} - \sqrt{\frac{5}{7}} \alpha \sqrt{\frac{5}{14}}= x.
	\label{append2}
\end{equation}
It is noted that the coefficient of the first term in (\ref{gamma4}), including $\Gamma_{7 \pm}^{(1)}$, becomes larger than that of the second term in (\ref{gamma4}), including $\Gamma_{7 \pm}^{(2)}$, if we use the values of (\ref{append4}).
However, it is allowable to discard the first term because the $\Gamma_{7 \pm}^{(1)}$ state is assumed to play a negligible role in forming the heavy fermion state as discussed above.
The normalization condition for the right part of eqs.(\ref{gamma5+})-(\ref{gamma5-}) requires $x=1$.
Nevertheless, by combining the normalization condition for $\Gamma_5^{(2)}$ symmetry in f$^2$ states, i.e., $\vert \alpha \vert^2 + \vert \beta \vert^2 = 1$, there are no solutions for these coefficients as far as $ 2\sqrt{11}/7 < x \le 1\ (2\sqrt{11}/7 \simeq 0.947 \cdots)$.
This is because we have discarded the states relating to $\Gamma_{7}^{(1)}$ as discussed above, and increased the weight of the remaining terms in eqs.(\ref{gamma5+})-(\ref{gamma5-}).
In view of such a situation, for simplicity, we use the pseudospin representations, $\vert \uparrow, \uparrow \rangle$ and $\vert \downarrow, \downarrow \rangle$, written in eqs.(\ref{2.2c})-(\ref{2.2d}), respectively, as $\Gamma_5^{(2)}$ states instead of using $\alpha$ and $\beta$.

\vspace{-4mm}
\begin{thebibliography}{99} 
\bibitem{lohneysen}
	H. L\"ohneysen, A. Rosch, M. Vojta, and P. W\"olfle: Rev. Mod. Phys. {\bf 79} (2007) 1015.
\bibitem{imada}
	M. Imada, A. Fujimori, and Y. Tokura: Rev. Mod. Phys. {\bf 70} (1998) 1039.
\bibitem{cox}
	D. L. Cox: Phys. Rev. Lett {\bf 59} (1987) 1240.
\bibitem{cox1}
	D. L. Cox and A. Zawadowski: Adv. Phys. {\bf 47} (1998) 599: and references therein.
\bibitem{yotsuhashi}
	S. Yotsuhashi, K. Miyake and H. Kusunose: J. Phys. Soc. Jpn. {\bf 71} (2002) 389.
\bibitem{hattori}
	K. Hattori and K. Miyake: J. Phys. Soc. Jpn {\bf 74} (2005) 2193.
\bibitem{kawae}
	T. Kawae, T. Yamamoto, K. Yurue, N. Tateiwa, K. Takeda, and T. Kitai: J. Phys. Soc. Jpn. {\bf 72} (2003) 2141.
\bibitem{wilson}
	K. G. Wilson: Rev. Mod. Phys. {\bf 47} (1975) 773.
\bibitem{krishna}
	H. R. Krishna-murthy, J. W. Wilkins and K. G. Wilson: Phys. Rev. B {\bf 21} (1980) 1003.
\bibitem{cragg}
	D. M. Cragg, P. Lloyd and P. Nozi$\grave{\rm e}$res: J. Phys. C {\bf 13} (1980) 803.
\bibitem{pang}
	H. B. Pang and D. L. Cox: Phys. Rev. B {\bf 44} (1991) 9454.
\bibitem{ott}
	H. R. Ott, H. Rudigier, Z. Fisk and J. L. Smith: Phys. Rev. Lett. {\bf 50} (1983) 1595.
\bibitem{ott2}
	H. R. Ott, H. Rudigier, E. Felder, Z. Fisk and J. L. Smith: Phys. Rev. B {\bf 33} (1986) 126.
\bibitem{gegenwart}
	P. Gegenwart, C. Langhammer, R. Helfrich, N. Oeschler, M. Land, J. S. Kim, G. R. Stewart, and F. Steglich: Physica. C {\bf 408-410} (2004) 157-160.
\bibitem{kim}
	J. S. Kim, B. Andraka, C. S. Jee, S. B. Roy, and G. R. Stewart: Phys. Rev. B {\bf 41} (1990) 11073.
\bibitem{kusunose}
	H. Kusunose and H. Ikeda: J. Phys. Soc. Jpn {\bf 74} (2005) 405.
\bibitem{ikeda}
	H. Ikeda and K. Miyake: J. Phys. Soc. Jpn {\bf 66} (1997) 3714.
\end{thebibliography}
\end{document}